\documentclass[aps,prb,twocolumn,amsmath,amsfonts,superscriptaddress]{revtex4-2}
\usepackage{graphicx}
\usepackage{dcolumn}
\usepackage{bm}
\usepackage{boxedminipage}
\usepackage{amsmath}
\usepackage{graphicx}
\usepackage{amsfonts}
\usepackage{feynmf}
\usepackage{mathrsfs}
\usepackage{listings}
\usepackage{algorithm}
\usepackage{algorithmic}
\usepackage{color}
\usepackage{braket}
\usepackage{hyperref,cleveref}
\usepackage{mhchem} 
\usepackage{ulem}

\newcommand{\nn}{\nonumber}
\newcommand{\beq}{\begin{eqnarray}}
\newcommand{\eeq}{\end{eqnarray}}

\newcommand{\CA}{\ce{Cd3As2}}

\begin{document}

\title{Intrinsic spin Nernst effect\\ in topological Dirac and magnetic Weyl semimetals}
\author{Taiki Matsushita}
\affiliation{Department of Physics, Graduate School of Science, Kyoto University, Kyoto 606-8502, Japan}
\affiliation{Center for Gravitational Physics and Quantum Information, Yukawa Institute for Theoretical Physics, Kyoto University, Kyoto 606-8502, Japan}
\author{Akihiro Ozawa}
\affiliation{Institute for Solid State Physics, The University of Tokyo, Kashiwa 277-8581, Japan}
\author{Yasufumi Araki}
\affiliation{Advanced Science Research Center, Japan Atomic Energy Agency, Tokai, Ibaraki 319-1195, Japan}
\author{Junji Fujimoto}
\affiliation{Department of Electrical Engineering, Electronics, and Applied Physics, Saitama University, Saitama, 338-8570, Japan}
\author{Masatoshi Sato}
\affiliation{Center for Gravitational Physics and Quantum Information, Yukawa Institute for Theoretical Physics, Kyoto University, Kyoto 606-8502, Japan}

\date{\today}

\begin{abstract}
We investigate the intrinsic spin Nernst effect (SNE), a transverse spin current induced by temperature gradients, in topological Dirac semimetals (TDSMs) and magnetic Weyl semimetals (MWSMs) with Ising spin-orbit coupling. The intrinsic SNE is described by the spin Berry curvature, which reflects the geometric nature of TDSMs and MWSMs. We clarified that the intrinsic SNE becomes significant when the Fermi energy is near, but slightly deviates from, the energy of the point nodes. In this situation, Bloch electrons with strong spin Berry curvature contribute to the SNE while avoiding carrier compensation between electrons and holes. 
We found that in TDSMs with small Fermi surfaces, the spin Nernst angle, which measures the efficiency of the SNE, is larger than that observed in heavy metals. This suggests that TDSMs with small Fermi surfaces can achieve efficient heat-to-spin current conversion. In MWSMs, variation in the magnitude of the exchange coupling with magnetic moments significantly changes the SNE, affecting both the direction and magnitude of the spin Nernst current.
This implies that ferromagnetic transitions can be used to reverse the spin Nernst current. These results provide the fundamentals for future topological spin caloritronics.
\end{abstract}

\maketitle

\section{Introduction}
The spin current is a central object in spintronics, enabling the control of magnetic orderings and textures.
In addition to electric current and voltage, heat current, which often arises as waste heat, has also been considered a source of spin current.
This field of study, known as spin caloritronics, aims for efficient heat-to-spin conversion and has thrived over the decades \cite{bauer2012spin}.
The spin Seebeck effect, a spin current induced by temperature gradients, has been widely discussed theoretically and experimentally~\cite{uchida2008observation, Adachi2011linear}.
This effect was first observed in the metallic ferromagnet permalloy (Ni$_{81}$Fe$_{19}$), and this discovery sparked significant interest in spin caloritronics~\cite{uchida2008observation}.
Subsequent research investigated the spin Seebeck effect in various materials, extending the concept to insulators, antiferromagnets, and superconductors~\cite{adachi2013theory,uchida2014longitudinal,wu2016antiferromagnetic,jaworski2010observation,uchida2010spin, PhysRevB.81.214418, PhysRevB.93.014425, PhysRevB.94.014427, PhysRevB.89.014416,jaworski2012giant,uchida2010observation, PhysRevLett.107.027205,umeda2018spin,shiomi2017oscillatory,sharma2023spin,matsushita2024spin}.

\begin{figure}[t]
\includegraphics[width=85mm]{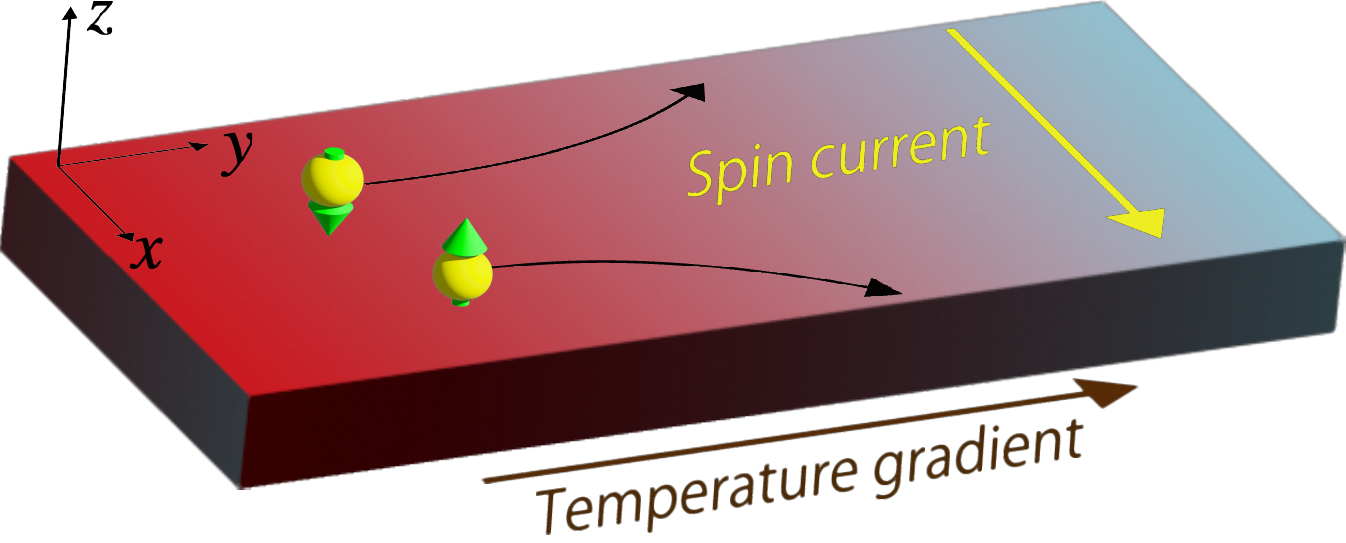}
\caption{
Schematic image of the SNE.
The brown and yellow arrows represent temperature gradients and the transverse spin current induced by the SNE, respectively.
The red and blue colors indicate the high and low temperatures.
}
\label{fig: image_SNE}
\end{figure}

The spin Nernst effect (SNE), a transverse spin current induced by temperature gradients, is one of the central phenomena in spin caloritronics (see Fig.~\ref{fig: image_SNE})~\cite{bauer2012spin}.
The SNE was observed in heavy metals such as tungsten and platinum~\cite {sheng2017,bose2018,bose2019}.
In contrast to the spin Seebeck effect, which is prohibited by time-reversal symmetry in many cases~\cite{Seemann2015, Wimmer2015},
the SNE enables the generation of the spin current with the temperature gradient in nonmagnetic materials.
Hence, the importance of the SNE has grown significantly.
However, theoretical research on the SNE is scant because a microscopic theory for the SNE was absent until recently~\cite{cheng2008,chuu2010,dyrdal2016,dyrdal2016a}.
Recently, a microscopic theory of the SNE was presented~\cite{fujimoto2024microscopic},
based on the Kubo-Luttinger theory~\cite{luttinger1964theory}.
This microscopic theory
satisfies desirable properties,
such as the spin-current version of the Mott formula~\cite{ashcroft}, which associates the SNE with the spin Hall effect (SHE) at low temperatures.

Nontrivial geometrical structures in momentum space can give rise to a large spin Nernst signal, and in certain cases, the SNE is directly linked to nontrivial topology in momentum space because the intrinsic contribution to the SNE is computed from the spin Berry curvature~\cite{PhysRevB.101.235201, Lau2023, fujimoto2024microscopic};
\beq
\label{eq: bsz}
b_{nz}^{s_z}=i\sum_{m(\neq n)}\frac{\langle u_n|J^{s_z}_x|u_m\rangle \langle u_m|v_y|u_n\rangle-{\rm c.c.}}{(\epsilon_n-\epsilon_m)^2},
\label{eq: SBC}
\eeq
where $J^{s_z}_x(\bm k)$ and $v_y(\bm k)$ are the spin current and velocity operators of the Bloch electrons, and $\epsilon_n(\bm k)$ is the eigenenergy of the Bloch state $|u_n(\bm k)\rangle$.
In the vicinity of band-touching points such as Dirac or Weyl points, the spin Berry curvature becomes divergently large, and its effects are significant.
Moreover, in spin-conserved systems, the spin Berry curvature reduces to the difference in the Berry curvature between the spin subspaces. In such cases, the spin Berry curvature is directly related to the band topology and is generated by Dirac or Weyl points~\cite{RevModPhys.82.1959, RevModPhys.90.015001}.

The above discussion implies that Dirac and Weyl semimetals with (approximate) spin conservation potentially exhibit a large spin Nernst signal, which reflects their nontrivial topology in momentum space.
Such systems include topological Dirac semimetals (TDSMs)~\cite{doi:10.1126/science.1245085,liu2014stable,neupane2014observation, araki2021long, PhysRevB.85.195320, young2012,PhysRevX.6.041021, burkov2016} and magnetic Weyl semimetals (MWSMs)~\cite{ozawa2019two,PhysRevB.97.235416,wang2018large,Guin2019zero,Yang2020,Noguchi2024,PhysRevApplied.21.014041} with Ising spin-orbit coupling (SOC).
The Ising SOC acts only on a single spin component, keeping the corresponding spin component preserved, and sometimes realizes the TDSM and MWSM states~\cite{PhysRevB.85.195320,ozawa2019two,Lau2023}.

In this paper, we investigate the intrinsic SNE in TDSMs and MWSMs with Ising SOC. We examine how the spin Nernst conductivity (SNC) depends on temperature, chemical potential, and exchange coupling with ferromagnetic moments to identify the conditions for a strong spin Nernst signal. We find that the intrinsic SNE becomes significant when the chemical potential is close to, but slightly deviates from, the energy of the point nodes. In such situation, Bloch electrons with strong spin Berry curvature contribute to the SNE while avoiding carrier compensation between electrons and hole carriers. 
We also show that TDSMs can achieve a large spin Nernst angle (SNA), which measures the efficiency of the SNE compared to that observed in heavy metals, when the Fermi surfaces are small. This large SNA reflects the nontrivial geometrical structure of TDSMs and enables efficient spin current generation with a temperature gradient. Furthermore, we find that in MWSMs, the variation in
the exchange coupling with the magnetic moments significantly changes the SNC, affecting both its magnitude and sign.

The rest of this paper is organized as follows.
We describe the theoretical formulation of the SNE in Sec.~\ref{sec2}.
In Sec.~\ref{sec3}, the model of TDSMs and MWSMs is presented.
Sections~\ref{sec4} and~\ref{sec5} constitute the main parts of this paper.
We discuss the SNE in TDSMs in Sec.~\ref{sec4} and in MWSMs in Sec.~\ref{sec5}.

\section{Theoretical Formulation}
\label{sec2}
\begin{figure*}
\includegraphics[width=160mm]{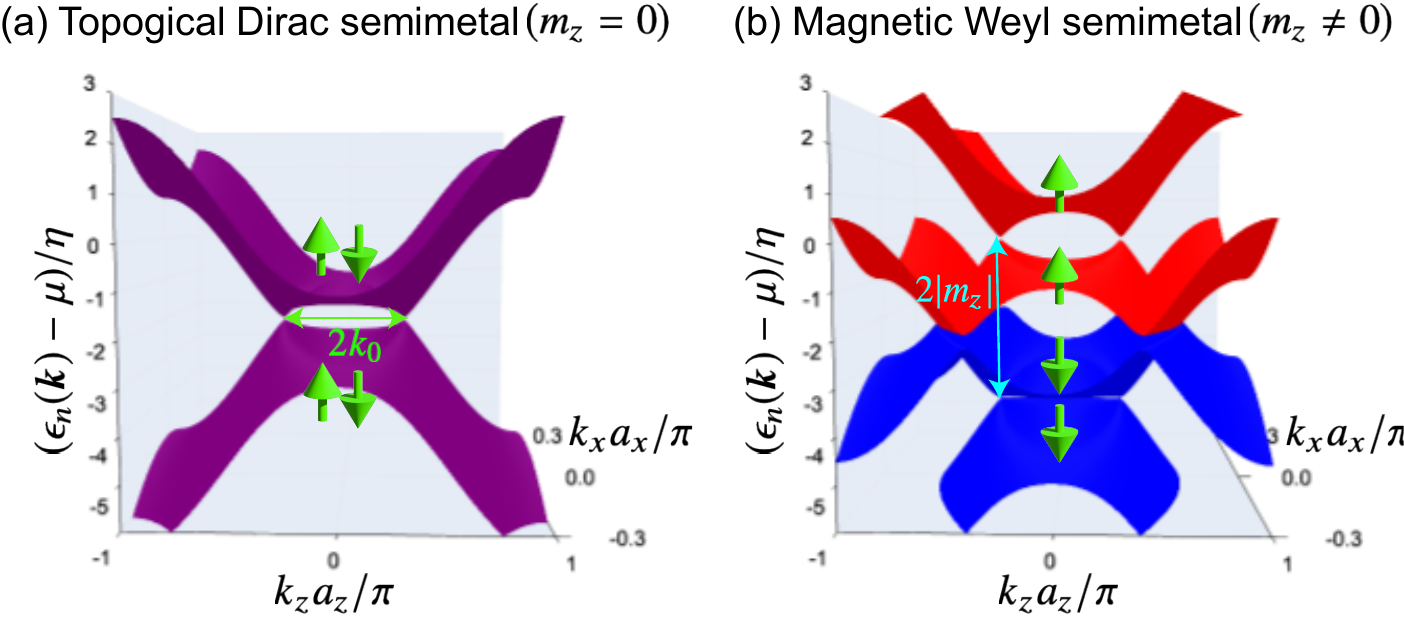}
\caption{Energy dispersion of (a) the TDSM state $(m_z=0)$ and (b) the MWSM state $(m_z\neq 0)$ computed from the $4\times 4$ tight binding Hamiltonian~\eqref{eq: Hamiltonian}.
}
\label{fig: energy band}
\end{figure*}

In this section, we explain the formulation of the SNE. 
The SNE arises from two mechanisms: intrinsic and extrinsic (impurity-induced)~\cite{cheng2008, dyrdal2016, dyrdal2016a, PhysRevLett.106.056601, PhysRevLett.109.026601, PhysRevB.87.075301, PhysRevLett.128.097001,fujimoto2024microscopic}.
In this paper, we focus on the intrinsic mechanism of the SNE, which is related to the geometric properties of three-dimensional point node semimetals.
We neglect the extrinsic mechanism because it is suppressed in Dirac and Weyl semimetal states with small Fermi surfaces~\cite{takahashi2008spin}.

Let us consider a temperature gradient applied along the $y$-direction ($-\partial_y T$) and a spin current flowing in the $x$-direction with spin polarization along the $z$-direction ($\langle J_x^{s_z}\rangle $), as shown in Fig.~\ref{fig: image_SNE}.
In the linear response regime, the transverse spin current is expressed as,
\beq
\label{eq: Einstein_spin}
\langle J_x^{s_z}\rangle &=&\alpha_{xy}^{{\rm (dir)}s_z}(-\partial_y T)+\sigma_{xy}^{s_z}E_y^{\rm ind}.
\eeq
Here, $J_x^{s_z}(\bm k)=\frac{1}{2}\{\frac{\hbar}{2}s_z, v_x(\bm k)\}$ is the spin current operator, where $\bm s=(s_x,s_y,s_z)$ represents the spin operator, and $\bm v(\bm k)=\partial H(\bm k)/\partial \bm k=(v_x(\bm k),v_y(\bm k),v_z(\bm k))$ is the velocity operator derived from the Hamiltonian $H(\bm k)$. The notation $\langle \cdots \rangle$ indicates the expectation value in the nonequilibrium state.

Equation \eqref{eq: Einstein_spin} shows the two contributions to the SNE.
The first term, which is proportional to $-\partial_y T$, represents the transverse spin current directly induced by the temperature gradient. We therefore refer to $\alpha_{xy}^{{\rm (dir)}s_z}$ as ``the direct contribution to the SNC" and it is expressed as~\cite{fujimoto2024microscopic},
\beq
\label{eq: SNC}
\alpha_{xy}^{{\rm (dir)}s_z}=\frac{1}{T}\sum_{n,\bm k}b_{nz}^{s_z}(\bm k) \int^\infty_{\epsilon_n(\bm k)-\mu}d\epsilon \ \epsilon \frac{\partial f(\epsilon)}{\partial \epsilon},
\eeq
where $f(\epsilon)=1/(e^{\beta \epsilon}+1)$ is the Fermi distribution function, $\epsilon_n(\bm k)$ is the energy of the Bloch state $|u_n(\bm k)\rangle$, and $\mu$ is the chemical potential. $b_{nz}^{s_z}(\bm k)$ is the spin Berry curvature defined by Eq.~\eqref{eq: bsz}~\cite{PhysRevB.101.235201}.
$E_y^{\rm ind}$ in the second term of Eq.~(\ref{eq: Einstein_spin}) represents the electric field along the $y$-direction induced by the Seebeck effect.
This electric field modifies the SNE through the SHE~\cite{PhysRevLett.109.026601,fujimoto2024microscopic}, which is described by the spin Hall conductivity (SHC)~\cite{RevModPhys.87.1213},
\beq
\label{eq: SHC}
\sigma_{xy}^{s_z}=-e\sum_{n,\bm k}b_{nz}^{s_z}(\bm k)f(\epsilon_n(\bm k)-\mu).
\eeq
To clarify this modification, we assume the open circuit condition in the $y$-direction, ensuring the absence of the electric current along the $y$-direction.
This condition requires that the Seebeck voltage compensates the electric current along the $y$-direction.
To describe this compensation, we write the electric current along the $y$-direction as,
\beq
\label{eq: Einstein_electric}
\langle J_y\rangle &=&\alpha_{yy}(-\partial_y T)+\sigma_{yy}E_y,
\eeq
and set $\langle J_y\rangle=0$, where $J_y(\bm k)=-e v_y(\bm k)$ is the electric current operator defined by the electron charge, $-e$.
$\alpha_{yy}$ and $\sigma_{yy}$ are the longitudinal thermoelectric and electric conductivities, respectively.
According to the Kubo formula, these conductivities are given by~\cite{mahan2013many,fujimoto2022seebeck},
\beq
\alpha_{yy}&=&\frac{e}{T}\sum_{\bm k}\int \frac{d\epsilon}{2\pi } \frac{\partial f(\epsilon)}{\partial \epsilon}{\rm Tr}\left[v_yG^{\rm R}\epsilon v_yG^{\rm A}\right],\\
\sigma_{yy}&=&-e^2\sum_{\bm k}\int \frac{d\epsilon}{2\pi } \frac{\partial f(\epsilon)}{\partial \epsilon}{\rm Tr}\left[v_yG^{\rm R}v_yG^{\rm A}\right],
\eeq
where $G^{\rm R}(\epsilon,\bm k)=G^{\rm A\dag}(\epsilon,\bm k)=1/(\epsilon+\mu-H(\bm k)+i\hbar/2\tau)$ is the retarded Green function with the relaxation time, $\tau$, and ${\rm Tr}$ repesents the trace of matrices in the spin and orbital spaces.
With these longitudinal conductivities, we obtain~\cite{mahan2013many,fujimoto2022seebeck},
\begin{subequations}
\beq
\label{eq: Seebeck}
E^{\rm ind}_y&=&-S_{yy}(-\partial_y T),\\
S_{yy}&=&\frac{\alpha_{yy}}{\sigma_{yy}}.
\eeq
\end{subequations}
Substituting Eq.~\eqref{eq: Seebeck} into Eq.~\eqref{eq: Einstein_spin}, we rearrange it as,
\begin{subequations}
\beq
\label{eq: Einstein_spin2}
\langle J_x^{s_z}\rangle &=& \alpha_{xy}^{s_z}(-\partial_y T),\\
\label{eq: total_SNC}
\alpha_{xy}^{s_z}&=&\alpha_{xy}^{{\rm (dir)} s_z}+\alpha_{xy}^{{\rm (indir)} s_z}.
\eeq
\end{subequations}
Equation~\eqref{eq: total_SNC} clearly shows the two contributions to the SNC.
We refer to the contribution from the Seebeck effect, $\alpha_{xy}^{{\rm (indir)} s_z}=-\sigma_{xy}^{s_z}S_{yy}$, as ``the indirect contribution to the SNC."
Note that the Seebeck voltage is insensitive to the relaxation time because both longitudinal thermoelectric and electric conductivities are proportional to the relaxation time~\cite{mahan2013many,fujimoto2022seebeck}.
Hence, the indirect contribution to the SNC is also independent of disorder and reflects the geometric nature of Bloch electrons.
Throughout this paper, the relaxation time $\tau$ is treated as a fixed parameter for the simplicity of calculations.

Both the direct and indirect contributions to the SNC are closely linked to the band topology when the spin ($s_z$) is conserved.  
This is because the spin Berry curvature reduces to the difference between the Berry curvatures of the spin subspaces ($s_z=\pm 1$) and is generated by Dirac or Weyl points~\cite{RevModPhys.82.1959}. 
If $s_z = \pm 1$ is a well-defined quantum number in a spin-$1/2$ system, the spin current operator can be written as $J_x^{s_z}=\frac{\hbar}{2}\left(v_x^{s_z=+1}-v_x^{s_z=-1}\right)$, where ${\bm v}^{s_z=\pm1}$ is the velocity operator for each spin subspace.  
By applying this relation to Eq.~(\ref{eq: bsz}), the spin Berry curvature $b_{nz}^{s_z}(\bm k)$ can be expressed as,
\beq
\label{eq: spin_Berry_curvature}
b_{nz}^{s_z}(\bm k)=\frac{\hbar}{2}\left[ b_{nz}^{s_z=+1}(\bm k)-b_{nz}^{s_z=-1}(\bm k)\right],
\eeq
where $b_{nz}^{s_z=\pm 1}(\bm k)$ is the Berry curvature for each spin subspace~\cite{RevModPhys.82.1959}.  
Therefore, when $s_z$ is conserved, the SNC can be decomposed into the anomalous Nernst conductivity (ANC), $\alpha_{xy}=\langle J_x\rangle/(-\partial_x T)$, for each spin subspace:
\beq
\label{eq:spin decomposition}
\alpha_{xy}^{s_z} = -\frac{\hbar}{2e}\left[ \alpha_{xy}^{s_z =+1}-\alpha_{xy}^{s_z =-1}\right],
\eeq
where $\alpha_{xy}^{s_z =\pm 1}$ is the ANC for each spin subspace. In Eq.~\eqref{eq:spin decomposition}, the coefficient $\hbar/2e$ converts the electric current into the spin current.

The above discussion suggests that Dirac and Weyl semimetals with (approximate) spin conservation reflect the band topology and can generate a large spin Nernst current.  
The (approximate) spin conservation sometimes coexists with Dirac and Weyl points when these point nodes are realized by the Ising SOC, which acts only on a single spin component, preserving this component ($s_z$ in the above discussion) while violating the conservation of the other two components ($s_x$ and $s_y$).  
The Ising SOC has been discussed such as in  the TDSM states of Cd$_3$As$_2$ and Na$_3$Bi, and in the MWSM states of Co$_3$Sn$_2$S$_2$~\cite{doi:10.1126/science.1245085,liu2014stable,neupane2014observation,ozawa2019two, PhysRevX.6.041021, PhysRevB.101.235201}. 
In these Dirac or Weyl materials, Ising SOC is largely dominant near the point nodes compared with other types of SOC that violate the conservation of $s_z$. In this regime, the relation \eqref{eq:spin decomposition} is approximately satisfied; that is, the spin Berry curvature $b_{nz}^{s_z}(\bm k)$ is largely unaffected by the violation of spin conservation.
In the following sections, we consider the intrinsic SNE in TDSMs and MWSMs with the Ising SOC, demonstrating how their geometric nature generates a large spin Nernst signal that reflects the nontrivial topology in momentum space.

\section{Model}
\label{sec3}
In this section, we introduce a minimal model for TDSMs and MWSMs with the Ising SOC.  
Dirac points have two-fold spin degeneracy, so the minimal model is described by a Hamiltonian in a $4 \times 4$-matrix form.
This minimal model includes two spin times two other internal degrees of freedom, such as orbitals or sublattices.  
When the system contains ferromagnetic moments,
the exchange coupling lifts the spin degeneracy at the Dirac points, causing a transition from the TDSM state to the spin-polarized MWSM state.  
In this paper, we consider a minimal model for TDSMs with two orbitals near the Fermi energy forming the Dirac points.  
The $4 \times 4$ tight-binding Hamiltonian, including the exchange coupling term for the MWSM state, is given by~\cite{PhysRevB.85.195320,yang2014},
\beq
\label{eq: Hamiltonian}
H(\bm k) &=& H_{\rm 0}(\bm k) + H_{\rm exc},\\
\label{eq: Hamiltonian_TDSM}
H_{\rm 0}(\bm k) &=& \left[M - t_{xy}(\cos k_x a_x + \cos k_y a_y) - t_z \cos k_z a_z \right] s_0 \sigma_z \nn \\
&+& \eta (\sin k_x a_x s_z \sigma_x - \sin k_y a_y s_z \sigma_y) \nn \\
&+& (\beta + \gamma) \sin k_z a_z (\cos k_y a_y - \cos k_x a_x) s_x \sigma_x \nn \\
&-& 2 (\beta - \gamma) \sin k_x a_x \sin k_y a_y \sin k_z a_z s_y \sigma_x,\\
\label{eq: Exc_cupling}
H_{\rm exc} &=& m_z s_z,
\eeq
where $\bm s = (s_x, s_y, s_z)$ and $\bm \sigma = (\sigma_x, \sigma_y, \sigma_z)$ are the Pauli matrices, and $s_0$ and $\sigma_0$ are the identity matrices for the spin and orbital spaces.  
$t_{xy}$ represents the hopping energy in the $x$- and $y$-directions, while $t_z$ represents the hopping energy in the $z$-direction.  
$M$ is the relative energy between the two orbitals near the Fermi energy, and $\eta$ represents the Ising SOC that preserves $s_z$.

\begin{figure*}
\includegraphics[width=180mm]{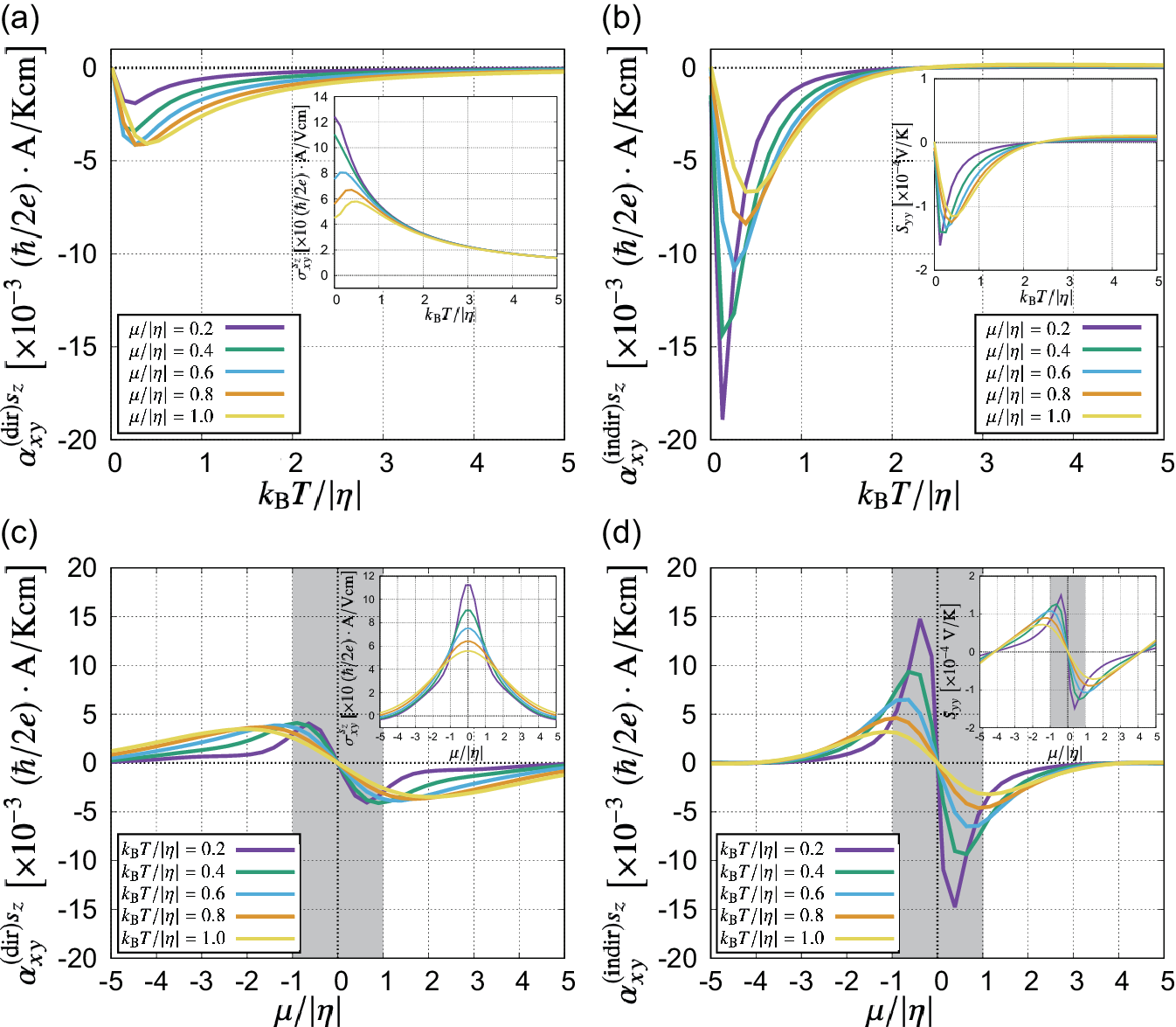}
\caption{
The temperature dependence of (a) the direct contribution and (b) the indirect contribution of the SNC in TDSMs with several values of the chemical potential.
The chemical potential dependence of (c) the direct contribution and (d) the indirect contribution of the SNC in TDSMs with the several values of the temperature.
The insets of panels (a) and (b) show the temperature dependence of the SHC and the Seebeck coefficient, respectively, while the insets of panels (c) and (d) show the chemical potential dependence of them.
The shaded area in panels (c) and (d) represents the energy region where the band structure becomes linear around the Dirac points.
For the calculations, we use the  $a_x=a_y=a_z=2$ nm\;  and $\eta=2.22\times 10^{-3} {\rm eV}$. 
The other parameters are set as $t_{xy}/\eta=1,\; t_z/\eta=1,\; M/\eta=4,\; \beta/\eta=0.2,\; \gamma/\eta=0.1$ and $\hbar/2\tau \eta=0.01$.
}.
\label{fig: Tdep_SNC_TDSM}
\end{figure*}

\begin{figure}
\includegraphics[width=85mm]{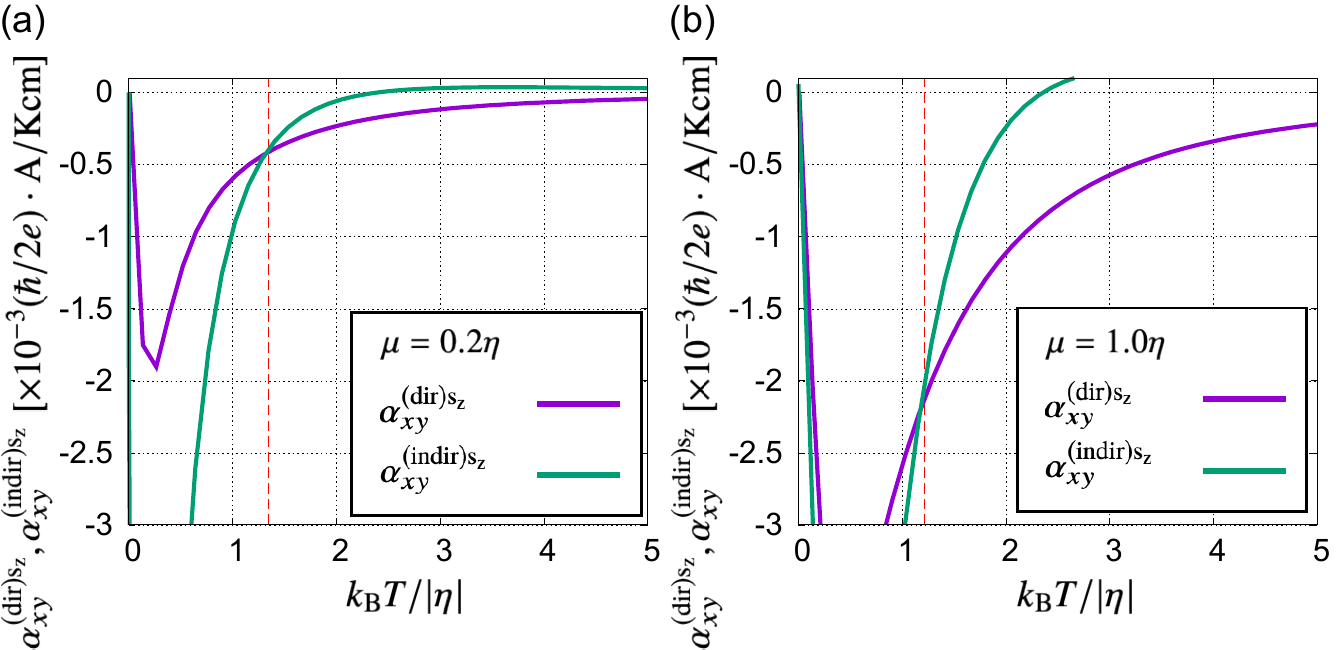}
\caption{
Comparison of the direct and indirect contributions to the SNC for (a) $\mu=0.2\eta$ and (b) $\mu=1.0\eta$. The other parameters are set the same as those in Fig.~\ref{fig: Tdep_SNC_TDSM}. The vertical red dashed lines represent the temperature $T_c$ at which the direct contribution surpasses the indirect one.
}
\label{fig: Tdep_SNC_mu0210}
\end{figure}
$H_{\rm 0}(\bm k)$ is the Hamiltonian for TDSMs, which respects the symmetries of time-reversal ($\mathcal{T}$), spatial inversion ($\mathcal{P}$), mirror reflection in the $xy$-plane ($\mathcal{M}_{xy}$), and four-fold rotation around the $z$-axis ($\mathcal{C}^z_4$).  
For $|M - 2t_{xy}| < |t_z|$, the low-energy band structure has two Dirac points at,
\beq
    \label{eq:Dirac-points}
    \bm k_0^{\lambda} = (0, 0, \lambda k_{\rm 0}) = \left(0, 0, \frac{\lambda}{a_z} \arccos\frac{M - 2t_{xy}}{t_z}\right),
\eeq
where $\lambda = \pm 1$.  
The bands are linearly dispersed around these Dirac points (see Fig.~\ref{fig: energy band} (a)).  
Up to the first order in ${\bm q} = {\bm k} - \bm k^{\lambda}_{\rm 0}$, the energy of the Bloch electrons can be linearized as,
\beq
&& \epsilon_n({\bm k}_0^{\lambda} + {\bm q}) \\
&& = \pm \sqrt{ (v_{{\rm D}x}^{\lambda} q_x)^2 + (v_{{\rm D}y}^{\lambda} q_y)^2 + (v_{{\rm D}z}^{\lambda} q_z)^2 } + O(q^2), \nonumber
\eeq
where ${\bm v}^{\lambda}_{\rm D} = (v_{{\rm D}x}^{\lambda}, v_{{\rm D}y}^{\lambda}, v_{{\rm D}z}^{\lambda}) = (\eta a_x, \eta a_y, \lambda t_z a_z \sin k_0 a_z)$ is the velocity of the Dirac fermions around $\bm k_0^{\lambda}$.

Dirac points are generally unstable and gapped out by perturbations.  
However, in TDSMs, Dirac points are protected by crystalline symmetries~\cite{yang2014}.
This protection makes the phase ``topological'', distinguishing it from Dirac semimetals without such protection.  
In this model, the Dirac points are protected by the four-fold rotational and $\mathcal{PT}$ symmetries.  
As long as these symmetries are maintained, the Dirac points remain stable even if perturbations are present~\cite{yang2014, PhysRevB.94.014510, PhysRevLett.115.187001}.

The third and fourth terms in Eq.~\eqref{eq: Hamiltonian_TDSM}, parametrized by the coefficients $\beta$ and $\gamma$, break the conservation of $s_z$.
These terms are of the order $\mathcal{O}({\bm q}^3)$, and hence the spin $s_z$ is largely conserved near the Dirac points.  
Because the intrinsic spin transport is primarily governed by the spin Berry curvature $b_{nz}^{s_z}(\bm k)$ around the Dirac points, the intrinsic SNE is mostly unaffected by the violation of the $s_z$-conservation due to the $\beta$- and $\gamma$- terms as long as the Fermi energy lies near the energy of the Dirac points.

$H_{\rm exc}$ describes the exchange coupling with the magnetic moments, where we assume the ferromagnetically ordered state along the $z$-direction.
The coefficient $m_z$ represents the energy of spin splitting.
This exchange coupling term breaks the $\mathcal{PT}$ symmetry and lifts the spin degeneracies at the Dirac points.
The lifting of spin degeneracies results in two pairs of Weyl points at the different energies separated by $2|m_z|$, as schematically shown in Fig.~\ref{fig: energy band} (b).
This band structure accounts for the ferromagnetic Weyl semimetals, where a pair of Weyl points lies near the Fermi energy.

Although the two-orbital tight-binding model described above was originally introduced to explain the band structures of Cd$_3$As$_2$ and Na$_3$Bi~\cite{PhysRevB.85.195320}, it can also be used to qualitatively describe the SNE in other Dirac or magnetic Weyl semimetals with Ising SOC, such as Co$_3$Sn$_2$S$_2$~\cite{ozawa2019two,Lau2023}, because this model consistently describes the distribution of the spin Berry curvature in Co$_3$Sn$_2$S$_2$ ~\cite{Lau2023}.

\begin{figure}[b]
\includegraphics[width=85mm]{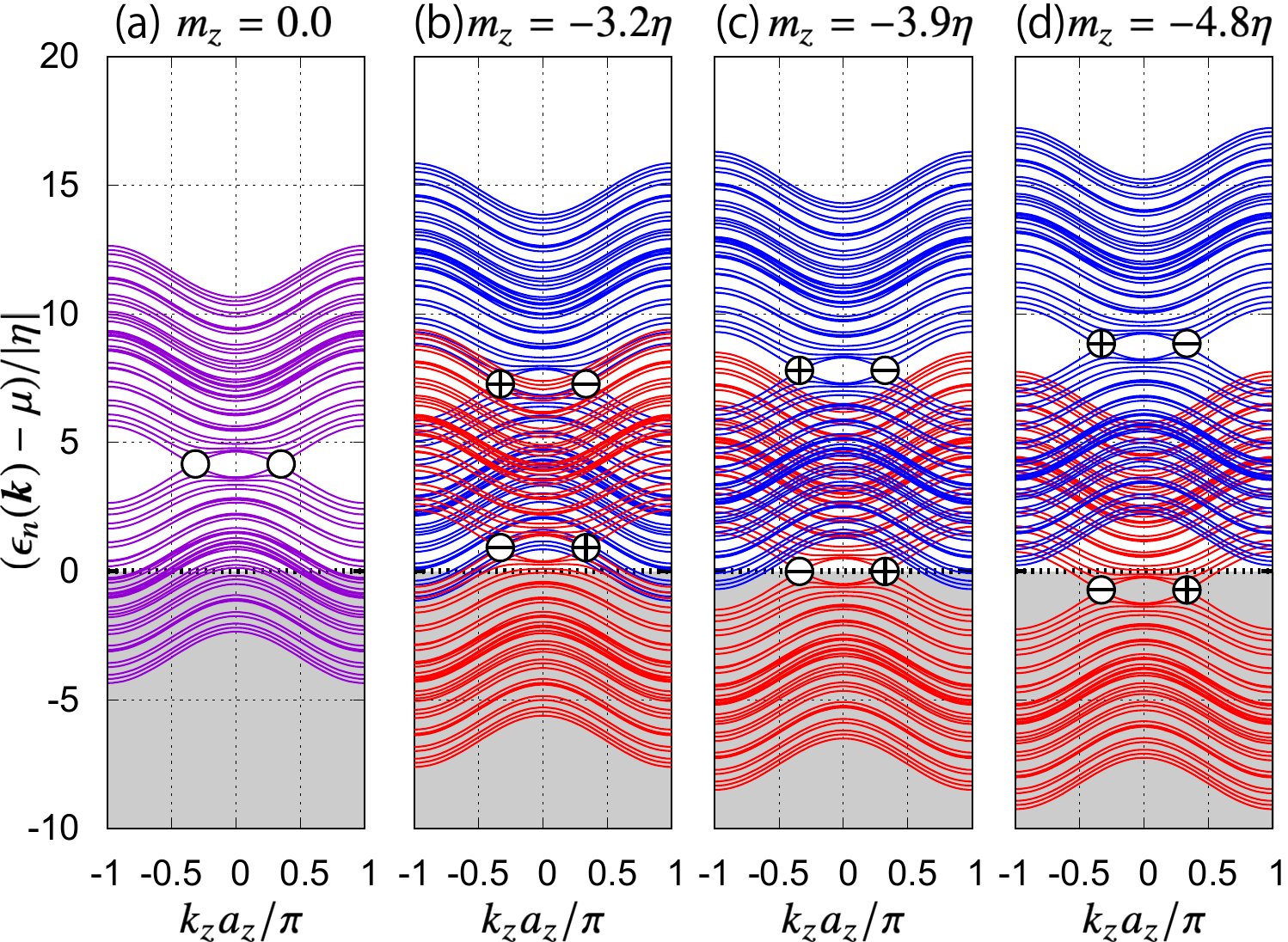}
\caption{
Energy spectra projected onto the $k_z$-axis for (a) the TDSM state ($m_z=0$) and (b-d) the MWSM states ($m_z\neq 0$) in the $s_z$-conserved systems ($\beta=\gamma=0$).
The purple curves represents the two-fold spin degeneracies, and the red and blue curves show the up-spin and down-spin states, respectively.
The symbols, $\bigcirc$ and $\oplus \;(\ominus)$ represent a Dirac point and a Weyl point with positive (negative) chirality.
The shaded area represents the energy region below the Fermi energy.
For visibility, the vertical axis is shared in all panels and the mesh size in the momentum space is set as $N_{k_x}\times N_{k_y} \times N_{k_z}=15 \times 15 \times 50$.
For the calculations, we set $N=1.02$ for all panels, $m_z=-3.2\eta$ for panel (b), $m_z=-3.9 \eta$ for panel (c), and $m_z=-4.8\eta$ for panel (d), respectively.
Other parameters are the same as those in Fig.~\ref{fig: Tdep_SNC_TDSM}.
} 
\label{fig: dispersion_MWSM}
\end{figure}

In this paper, all calculations were conducted with $\eta = 2.22$ meV and $a_x = a_y = 2$ nm as parameters for \CA~\cite{PhysRevB.85.195320}. 
The other parameters can be found in the captions of each figure.
We also set the mesh size in momentum space as $N_{k_x} \times N_{k_y} \times N_{k_z} = 200 \times 200 \times 200$ (except for the energy spectra shown in Fig.~\ref{fig: dispersion_MWSM}).
 
\section{Spin Nernst effect in topological Dirac semimetal state}
\label{sec4}
We have introduced the minimal model for the TDSMs with the Ising SOC.  
Here, we show that in TDSMs $(m_z=0)$, the SNE is sensitive to the energy of the Dirac points and is amplified when the energy of the Dirac points is close to, but not exactly at, the Fermi energy.  
Furthermore, we demonstrate that TDSMs can achieve a SNA larger than that observed in heavy metals.

Figure~\ref{fig: Tdep_SNC_TDSM} (a-b) shows the temperature dependence of the SNC, where panels (a) and (b) describes the direct and indirect contributions to the SNE, respectively.  
Both contributions vanish at zero temperature and are suppressed as \(k_{\rm B}T \rightarrow \infty\).  
At zero temperature, the SNE must vanish according to the thermodynamic principles.  
The SNE is subject to the compensation of electron and hole carriers similarly to the thermoelectric current~\cite{PhysRevB.11.573, PhysRevX.12.011037,fujimoto2022seebeck}.
As \(k_{\rm B}T \rightarrow \infty\), the carrier compensation becomes complete and the SNE vanishes.  
As shown in Fig.~\ref{fig: Tdep_SNC_TDSM} (a-b), the magnitudes of both contributions peak at an intermediate temperature and reach their maxima at \(k_{\rm B}T\sim \mathcal{O}(0.1\eta)\).

It depends on the temperature which of the two contributions, direct or indirect, surpasses the other.
At low temperature $k_{\rm B}T \lesssim \eta$, we find that the indirect contribution becomes significantly larger than the direct contribution.
On the other hand, at higher temperature, the indirect contribution becomes less dominant compared to the direct contribution due to the suppression of the Seebeck effect (see the inset of the panel (b)).
As shown in Fig.~\ref{fig: Tdep_SNC_mu0210},
the direct contribution surpasses the indirect contribution beyond the temperature, $T_c= 1.57 |\eta|/k_{\rm B} \sim 40.4{\rm K}$ for $\mu=0.2\eta$, and $T_c= 1.31 |\eta|/k_{\rm B} \sim 33.7{\rm K}$ for $\mu=\eta$.

Figure~\ref{fig: Tdep_SNC_TDSM} (c-d) shows how the SNC depends on the chemical potential, again split into (c) the direct contribution and (d) the indirect contribution of the SNC.
Both contributions vanish at $\mu=0$ and get suppressed as $|\mu| \rightarrow \infty$.
At $\mu = 0$, the Fermi energy lies exactly at the energy of the Dirac points, leading to complete carrier compensation between electrons and holes due to the particle-hole symmetric spectrum~\cite{PhysRevB.11.573, PhysRevX.12.011037,fujimoto2022seebeck}.  
In contrast, in the limit $|\mu| \rightarrow \infty$, the spin Berry curvature effect vanishes because the Fermi level becomes far away from the energy of the Dirac points,
resulting in the suppression of the SNE.
The magnitudes of both contributions peak at the intermediate chemical potential, reaching their maxima at  $|\mu| \sim \mathcal{O}(\eta)$.
This occurs because, at this point, Bloch electrons near the region with the strong spin Berry curvature are excited, amplifying the SNE while avoiding the carrier compensation.

At low temperature, the SNE is associated with the SHE and its direct contribution satisfies the spin current version of the Mott formula, which is given by~\cite{fujimoto2024microscopic},
\beq
\alpha_{xy}^{{\rm (dir)}s_z}(T,\mu)=\frac{\pi^2 T}{3}\frac{\partial \sigma_{xy}^{s_z}(T=0,\mu)}{\partial \mu}.
\eeq
As shown in Fig.~\ref{fig: Tdep_SNC_TDSM} (c), the SHC reaches a maximum at $\mu=0$ whereas the direct contribution of the SNC changes its sign at $\mu=0$.
This behavior is consistent with the spin current version of the Mott formula.

\begin{figure*}[t]
\includegraphics[width=180mm]{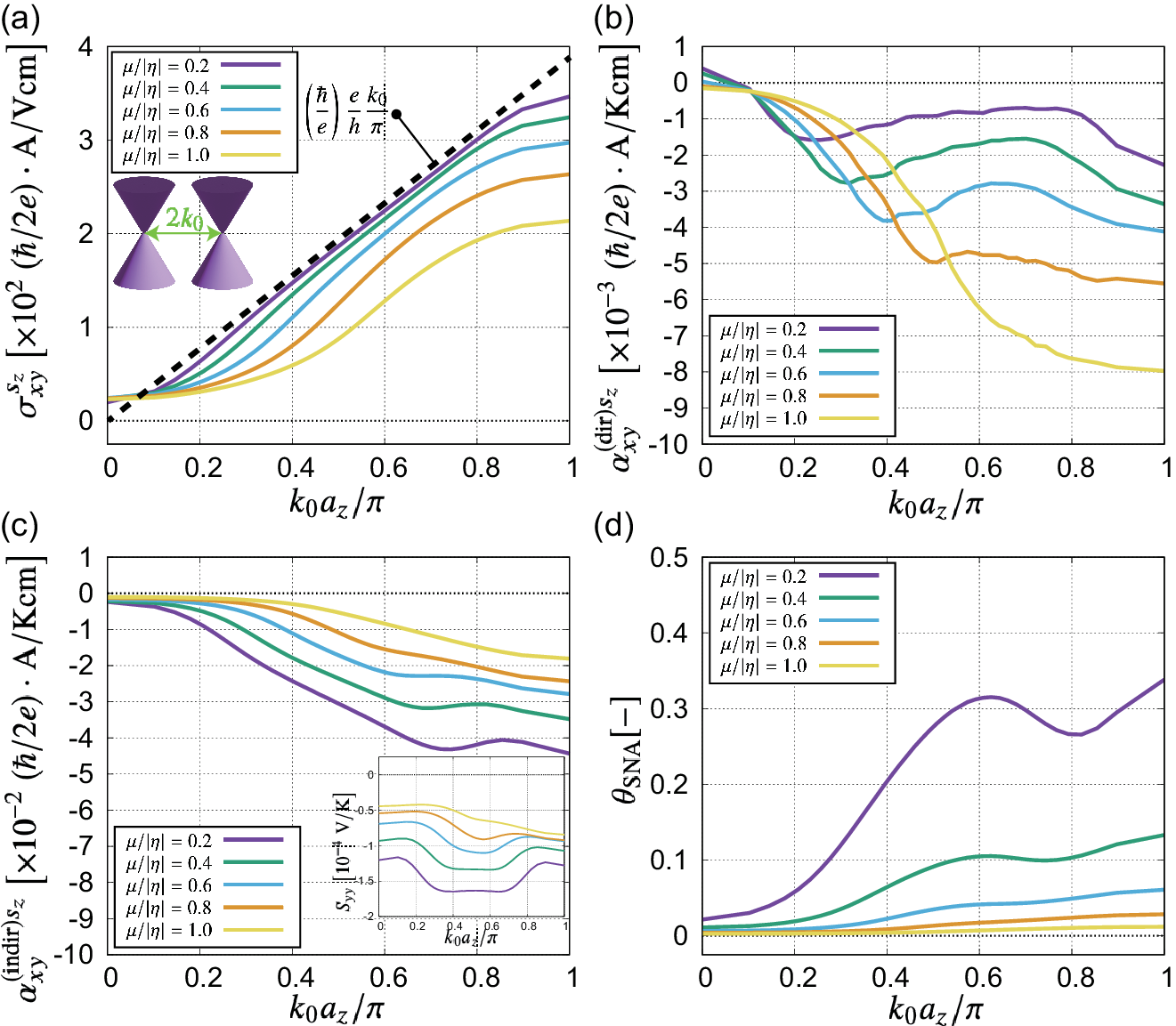}
\caption{
The $k_0$-dependences of (a) the SHC, (b) the direct contribution, (c) the indirect contribution of the SNC, and (d) the SNA in TDSMs with the several values of the chemical potential $\mu$.
$M$ is tuned to vary the positions of the Dirac points $(0,0,\pm k_0)$ on the $k_z$-axis. 
In panel (a), the configuration of the Dirac points is schematically explained, and the dotted black line represents the analytical relation in Eq.~\eqref{eq: QSHE}.
The inset of the panel (c) shows the $k_0$-dependences of the Seebeck coefficient.
For the calculation,s we set  $k_{\rm B}T = 0.1\eta$ and the other parameters the same as those in Fig.~\ref{fig: Tdep_SNC_TDSM}.
}
\label{fig: kDdep_SNC_TDSM}
\end{figure*}

Let us next discuss how the SHC and the SNC depend on the configurations of the Dirac points, ${\bm k}_0^{\lambda} = (0,0,\lambda k_0)$ (see the inset of  Fig.~\ref{fig: kDdep_SNC_TDSM} (a)).
Figure~\ref{fig: kDdep_SNC_TDSM} shows the calculated results for (a) the SHC, (b) the direct contribution, and (c) the indirect contribution of the SNC.
When $k_{\rm B}T=\mu=\beta=\gamma=0$, the SHC becomes proportional to $k_0$ and is expressed as~\cite{PhysRevB.101.235201,burkov2016},
\beq
\label{eq: QSHE}
\sigma_{xy}^{s_z}={\rm sign}(t_z)\left(\frac{\hbar}{e}\right)\frac{e}{h}\frac{k_0}{\pi}.
\eeq
This relationship is understood as follows.
When $\beta=\gamma=0$, $s_z$ is exactly conserved, allowing the spin-resolved description.
In this limit, the Hamiltonian of TDSM can be viewed as a family of the two-dimensional quantum spin Hall insulators parametrized by $k_z$ \cite{PhysRevLett.95.226801, PhysRevLett.95.146802}.
These two-dimensional systems are characterized by the spin Chern number,
\beq
{\rm Ch_s}(k_z)=\sum_{n\in {\rm occupied\;band}}\int \frac{dk_xdk_y}{4\pi} b_{nz}^{s_z}(\bm k),
\eeq
which becomes ${\rm Ch_s}(k_z)={\rm sign}(t_z)\Theta(k_0^2-k_z^2)$~\cite{PhysRevB.101.235201}, where $\Theta(x)$ is the step function.
The spin Chern number is a topological invariant, which takes an integer value at each $k_z$, associated with the quantized spin Hall conductivity in the two-dimensional systems as $\sigma_{xy}^{s_z}(k_z)=\left(\frac{\hbar}{e}\right)\frac{e}{h}{\rm Ch_s}(k_z)$~\cite{PhysRevLett.95.226801, PhysRevLett.95.146802}.
Summing up $\sigma_{xy}^{s_z}(k_z)$ over $k_z$, we obtain the SHC in three dimensions, resulting in Eq.~\eqref{eq: QSHE}.
The SHC deviates from Eq.~\eqref{eq: QSHE} for the finite Fermi energy because the occupied states in the conduction band partially cancel the contribution from the valence band.
As shown in Fig.~\ref{fig: kDdep_SNC_TDSM} (c), the indirect contribution of the SNC has a similar $k_0$-dependence to the SHC because the indirect contribution of the SNC is proportional to the SHC.

In contrast to the SHC, the direct contribution to the SNC is not proportional to $k_{\rm 0}$ (see Fig.~\ref{fig: kDdep_SNC_TDSM} (b)).
This is because only the thermally excited Bloch electrons near the Fermi energy contribute to the SNE, in contrast to the SHE for which all the occupied states below the Fermi energy contribute.
Specifically, we find that $\alpha_{xy}^{{\rm (dir)}s_z}$ tends to reach a saturation value for $|v_{{\rm D}z}^{\lambda}k_0| \gtrsim |\mu|$.
This is because, in this parameter region, the Fermi surfaces around the two Dirac points are well separated, which makes the geometrical structures of the carriers around the Fermi surfaces almost insensitive to the value of $k_0$.
Within the region $|v_{{\rm D}z}^{\lambda}k_0|\lesssim |\mu|$, where the two Dirac points are not well separated, the increase of $k_0$ enhances the direct contribution to the SNE.
This enhancement occurs because the increase of $k_0$ raises the number of the thermally excited Bloch electrons in the region $|k_z|<k_0$, where the spin Berry curvature influences the Bloch electrons.

With the SNCs calculated above,
we now evaluate the efficiency of the SNE in TDSMs.
A qualitative measure of the efficiency of the SNE is the SNA, which is defined as the ratio of the SNC to the longitudinal thermoelectric conductivity $\alpha_{yy}$~\cite{PhysRevLett.104.186403},
\beq
\theta_{\rm SNA}=\frac{2e}{\hbar}\frac{\alpha_{xy}^{s_z}}{\alpha_{yy}}.
\eeq
Our calculation result in Fig.~\ref{fig: kDdep_SNC_TDSM}(d) shows that the SNA in TDSMs reaches $\theta_{\rm SNA} \simeq 0.33$ at $\mu = 0.2$. This value is larger than the SNA measured in several heavy metals, such as tungsten ($|\theta_{\rm SNA}| = 0.14 \text{-} 0.22$)~\cite{sheng2017} and platinum ($|\theta_{\rm SNA}| = 0.20$)~\cite{bose2018,bose2019}. The SNA decreases as $\mu$ increases, due to the increase in $\alpha_{yy}$ from the large Fermi surfaces. These results suggest that the large SNA requires both a nontrivial geometric structure and low carrier density. We expect that TDSMs satisfying these conditions can be used to achieve highly efficient heat-to-spin current conversion.

\begin{figure}[t]
\includegraphics[width=80mm]{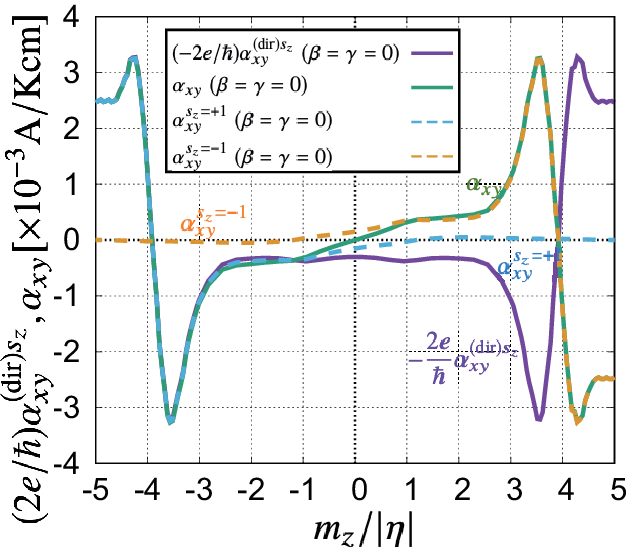}
\caption{
The $m_z$-dependences of the SNC, the ANC, and the ANC in each spin subspace in the $s_z$-conserved system $(\beta=\gamma=0)$.
We set $k_{\rm B}T=0.1\eta$ and $N=1.02$.
The other parameters are set as the same as those in Fig.~\ref{fig: Tdep_SNC_TDSM}.
}
\label{fig: mzdep_ANE_WM}
\end{figure}

\begin{figure*}[t]
\includegraphics[width=180mm]{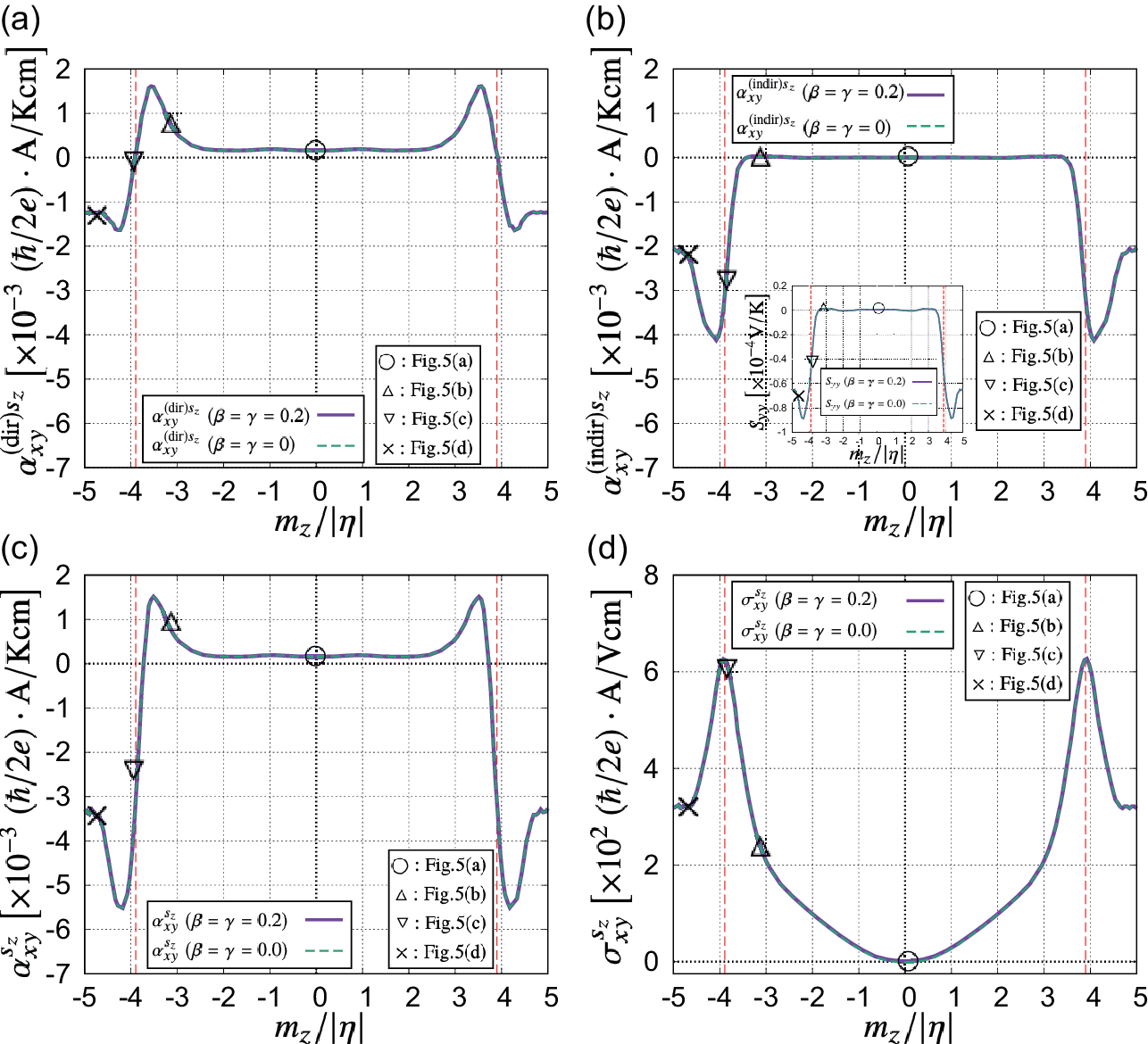}
\caption{
The $m_z$-dependences of (a) the direct contribution to the SNC, (b) the indirect contribution to the SNC, (c) the total SNC, and (d) the SHC in the $s_z$-conserved ($\beta=\gamma=0$) and nonconserved systems ($\beta=\gamma=0.2$). For all panels, the 
vertical red dashed lines represent the exchange coupling satisfying $|m_z|=|m_{zc}|$, and the symbols ($\bigcirc, \bigtriangleup, \bigtriangledown, \times$) indicate the correspondence to panels (a-d) in Fig.~\ref{fig: dispersion_MWSM}. The inset of panel (b) shows the $m_z$-dependence of the Seebeck coefficient. For all panels, we set $k_{\rm B}T=0.1\eta$ and $N=1.02$. The other parameters are the same as those in Fig.~\ref{fig: Tdep_SNC_TDSM}.
}
\label{fig: mzdep_SNE_WM}
\end{figure*}

\section{Spin Nernst effect in magnetic Weyl semimetal state}
\label{sec5}
So far, we have discussed the SNE in nonmagnetic TDSMs ($m_z=0$).
As described in Sec.~\ref{sec3}, our model can describe MWSMs with the Ising SOC by introducing exchange coupling with ferromagnetic moments ($m_z\neq 0$).
The magnetic moments align ferromagnetically below the Curie temperature, with $m_z\neq 0$, splitting a Dirac point into two Weyl points to exhibit the MWSM state.
To understand the effect of the ferromagnetic transition on the SNE, we next examine the SNE in the MWSM state ($m_z\neq 0$).
In Sec.~\ref{sec5}, we show that the variations in the spin splitting $m_z$ drastically change the SNC, affecting both its magnitude and sign.

Because the spin splitting by $m_z$ significantly alters the overall band structure, the chemical potential $\mu$ becomes highly sensitive to the spin-splitting energy (see Fig.~\ref{fig: dispersion_MWSM}). 
Therefore, instead of the chemical potential, we treat the electron density (or the electron number at each site) 
\begin{equation}
\label{eq:density}
N = \frac{1}{V} \sum_{n, \bm{k}} \Theta(\mu - \epsilon_n(\bm{k})),
\end{equation}
as the parameter for the calculations herein. 
Our model has four bands, and $N = 2$ corresponds to half-filling. 
Because the particle-hole symmetric spectrum prevents a spin current response to temperature gradients as well as a thermoelectric current response, we consider the hole-doped MWSM state with $N = 1.02$.

The exchange coupling with ferromagnetic moments lifts the spin degeneracy of Dirac points, turning them into two pairs of Weyl points. 
Figure~\ref{fig: dispersion_MWSM} shows the bulk energy spectra measured from the chemical potential for the case $\beta = \gamma = 0$, which allows for a spin-resolved analysis. 
These spectra are calculated for several values of $m_z$ and are projected onto the $k_z$ axis. 
Figure~\ref{fig: dispersion_MWSM}(a) corresponds to the nonmagnetic state $(m_z = 0)$ with two-fold spin degeneracy at each $\bm{k}$-point, while Figs.~\ref{fig: dispersion_MWSM}(b-d) show the energy spectra of the MWSM states $(m_z \neq 0)$ for different values of $m_z$. 

Figure~\ref{fig: dispersion_MWSM} shows that the energies of the point nodes, measured from the chemical potential, are sensitive to $m_z$. As described above, we treat the electron density as the given parameter for calculations instead of the chemical potential. The chemical potential $\mu$ is determined by solving Eq.~\eqref{eq:density} for a given electron density ($N = 1.02$) and $m_z$. In the nonmagnetic state $(m_z = 0)$, the Dirac points are located far above the Fermi energy (see Fig.~\ref{fig: dispersion_MWSM}(a)). When the exchange coupling term is introduced $(m_z \neq 0)$, it shifts the energy of one pair of Weyl points away from the Fermi energy, while the other pair moves closer to it (see Figs.~\ref{fig: dispersion_MWSM}(b-d)). The Weyl points that move away from the Fermi energy are almost irrelevant to the SNE because only the electrons near the Fermi energy contribute to thermal responses. For $m_z = -3.2 \eta$, the energy of the Weyl points for spin-up electrons is above the Fermi energy (see Fig.~\ref{fig: dispersion_MWSM}(b)), whereas for $m_z = -4.8\eta$, the Weyl points are below the Fermi energy (see Fig.~\ref{fig: dispersion_MWSM}(d)).
The energy of the Weyl points for spin-up electrons lies exactly on the Fermi energy when $m_z = -|m_{zc}| \equiv -3.9 \eta$ (see Fig.~\ref{fig: dispersion_MWSM}(c)). 
Hence, the energy of a pair of Weyl points is close to the Fermi energy for $|m_z| \sim |m_{zc}|$, and exactly at it when $|m_z| = |m_{zc}|$. This energy is above (below) the Fermi level for $|m_z| < |m_{zc}|$ ($|m_z| > |m_{zc}|$).

We now examine how the SNE depends on the exchange coupling, particularly considering the sign and magnitude of $m_z$. We first show that the sign of $m_z$ does not affect the spin Nernst current.
Figure~\ref{fig: mzdep_ANE_WM} illustrates the $m_z$ dependence of the direct contribution to the SNC and the ANC in the $s_z$-conserved system~($\beta=\gamma=0$), which allows us to distinguish contributions from each spin subspace (see Fig.~\ref{fig: mzdep_SNE_WM}(b) for the indirect contribution to the SNC). The change in the sign of $m_z$ corresponds to the reversal of the ferromagnetic moments, effectively transforming the system into its time-reversed counterpart. Figure~\ref{fig: mzdep_ANE_WM} clearly shows that the sign reversal of $m_z$ preserves the direction of the spin Nernst current while it reverses the sign of the ANC. This occurs because the spin current is even $(\mathcal{T}^{-1}J_x^{s_z}\mathcal{T}=J_x^{s_z})$, whereas the electric current is odd $(\mathcal{T}^{-1}J_x\mathcal{T}=-J_x)$ under time-reversal operation, $\mathcal{T}$. We also note that, in MWSMs with Ising SOC, the SNE is insensitive to the violation of $s_z$ conservation (see Figs.~\ref{fig: mzdep_SNE_WM} (a-c)).

We then show that the spin Nernst current is highly sensitive to the magnitude of the exchange coupling $|m_z|$. As shown in Figs.~\ref{fig: mzdep_SNE_WM}(a-b), all contributions to the SNC become large near $|m_z| = |m_{zc}|$, where the energy of a pair of Weyl points resides near the Fermi energy.
This is because Bloch electrons with strong spin Berry curvature contribute to the SNE. In addition, the direct contribution to the SNC changes its sign at $|m_z| = |m_{zc}|$ due to the sign change in the spin Berry curvature of conduction electrons.
In this calculation, the spin current reversal of the direct SNE causes the reversal of the total spin Nernst current (see Fig.~\ref{fig: mzdep_SNE_WM}(c)). The indirect contribution to the SNC does not change its sign (see Fig.~\ref{fig: mzdep_SNE_WM}(b)), because the Fermi surfaces from the Weyl bands away from the Fermi energy lead to incomplete carrier compensation at $|m_z| = |m_{zc}|$. Additionally, we observe that the SHC peaks at $|m_z| = |m_{zc}|$ and decreases as $|m_z|$ deviates from $|m_{zc}|$ (see Fig.\ref{fig: mzdep_SNE_WM}(d)).

\section{Conclusion}
In this paper, we established the intrinsic SNE in TDSMs and MWSMs as their nontrivial geometrical nature. The SNC has the direct and indirect contributions. The indirect contribution arises from the combination of the SHE and the Seebeck effect. We first investigated the relative magnitudes of these contributions in TDSMs. Both contributions vanish at zero temperature, are suppressed as $k_{\rm B}T \to \infty$, and peak at intermediate temperatures. At low temperatures, the indirect contribution becomes dominant. In contrast, as increasing temperature, the indirect contribution is more sharply suppressed, because it relies on the Seebeck effect. Therefore, the direct contribution exceeds the indirect contribution above a certain temperature, which is estimated to be $T \sim 33.7$ K for \CA.

We next examined the conditions for the strong spin Nernst signal. We found that the SNC becomes large when the energy of the Dirac points is close to, but slightly different from, the Fermi energy. In this situation, Bloch electrons with strong spin Berry curvature contribute to the SNE while avoiding carrier compensation between electrons and holes. We obtain the SNA larger than that observed in heavy metals~\cite{sheng2017,bose2018}, when the Fermi surfaces are small. This suggests that TDSMs with small Fermi surfaces can achieve highly efficient heat-to-spin current conversion.

Finally, we introduced the exchange coupling with ferromagnetic moments to understand the effect of the transition from the nonmagnetic TDSM state to the MWSM state. We found that the direction of the spin Nernst current is unaffected by the reversal of the ferromagnetic moments, because the spin current is even under the time-reversal operation. We also found that the SNE in MWSMs is sensitive to the magnitude of the exchange coupling. In MWSMs, the SNC becomes large when the energy of a pair of Weyl points resides near the Fermi energy, allowing the Bloch electrons with strong spin Berry curvature to contribute to the SNE. Besides this, we found that the spin Nernst current reverses when the exchange coupling pushes the energy of the Weyl points to the Fermi energy. This result implies that ferromagnetic phase transitions can be used to induce a reversal of the spin Nernst current in MWSMs, such as Co$_3$Sn$_2$S$_2$~\cite{ozawa2019two}.

The SHE, ANE, and anomalous Hall effect have been observed in Dirac and Weyl semimetals so far~\cite{wang2018large, ikeda2018anomalous, higo2018anomalous, liu2017anomalous, Guin2019zero, Yang2020, Lau2023, Noguchi2024}. 
Even though the SNE in these systems has not been explored yet, we expect that its future experimental detection is feasible as a new manifestation of their geometric nature.

\section*{Acknowledgement}
TM and MS was supported by JST CREST Grant No.~JPMJCR19T2.
TM was supported by JSPS KAKENHI Grant No.~JP24KJ0130.
YA was supported by JSPS KAKENHI, Grant No.~JP22K03538.
JF was supported by JSPS KAKENHI, Grant No.~JP22K13997.
MS was supported by JSPS KAKENHI, Grant No.~JP24K00569.
The numerical calculations were performed with computational facilities at YITP in Kyoto University.

\bibliography{ISNE.bib}
\bibliographystyle{apsrev}
\end{document}